\documentclass[11pt]{article}
\usepackage{graphics,epsfig}
\usepackage{amsmath,amssymb,amsthm}
\usepackage{url}

\newcommand{\bibnodot}[1]{}

\begin{document}

\title{Quantum Structure in Economics: The Ellsberg Paradox}

\author{Diederik Aerts and Sandro Sozzo\\
        \normalsize\itshape
        Center Leo Apostel for Interdisciplinary Studies and Department of Mathematics \\
        \normalsize\itshape
        Brussels Free University, Pleinlaan 2, 1050 Brussels, 
       Belgium \\
        \normalsize
        E-Mails: \textsf{diraerts@vub.ac.be, ssozzo@vub.ac.be}}
\date{}
\maketitle

\begin{abstract}
\noindent The \emph{expected utility hypothesis} and \emph{Savage's Sure-Thing Principle} are violated in real life decisions, as shown by the \emph{Allais} and \emph{Ellsberg paradoxes}. The popular explanation in terms of \emph{ambiguity aversion} is not completely accepted. As a consequence, uncertainty is still problematical in economics. To overcome these difficulties a distinction between \emph{risk} and \emph{ambiguity} has been introduced which depends on the existence of a Kolmogorovian probabilistic structure modeling these uncertainties. On the other hand, evidence of everyday life suggests that context plays a fundamental role in human decisions under uncertainty. Moreover, it is well known from physics that any probabilistic structure modeling contextual interactions between entities structurally needs a non-Kolmogorovian framework admitting a quantum-like representation. For this reason, we have recently introduced a notion of \emph{contextual risk} to mathematically capture situations in which ambiguity occurs. We prove in this paper that the contextual risk approach can be applied to the Ellsberg paradox, and elaborate a sphere model within our \emph{hidden measurement formalism} which reveals that it is the overall conceptual landscape that is responsible of the disagreement between actual human decisions and the predictions of expected utility theory, which generates the paradox. This result points to the presence of a \emph{quantum conceptual layer} in human thought which is superposed to the usually assumed \emph{classical logical layer}, and conceptually supports the thesis of several authors suggesting the presence of quantum structure in economics and decision theory. 
\end{abstract}

%%%%%%%%%%%%%%%%%%%%%%%%%%%%%%%%%%%%%%%%%%%%
%% MAINMATTER
%%%%%%%%%%%%%%%%%%%%%%%%%%%%%%%%%%%%%%%%%%%%

\section{The Sure-Thing Principle and the Ellsberg Paradox\label{ellsberg}}
The \emph{expected utility hypothesis} requires that in uncertain circumstances individuals choose in such a way that they maximize the expected value of `satisfaction' or `utility'. This hypothesis is the prevailing model of choice under uncertainty in economics, and is founded on the \emph{von Neumann-Morgenstern utility theory} \cite{vonneumannmorgenstern1944}. These authors provided a set of  `reasonable' axioms under which the expected utility hypothesis holds. One of the proposed axioms is the \emph{independence axiom} which expresses Savage's \emph{Sure-Thing Principle} \cite{savage1954}. Examples exist in the literature which show an inconsistency with the predictions of the expected utility hypothesis, namely a violation of the Sure-Thing Principle. These deviations are known as \emph{Allais} \cite{allais1953} and \emph{Ellsberg} \cite{ellsberg1961} \emph{paradoxes}. They at first sight indicate the existence of an \emph{ambiguity aversion}, that is, individuals prefer `sure' over `uncertain' choices. Several approaches have been forwarded to explain the deviations predicted by these paradoxes but none of them is completely satisfying.

Savage introduced the Sure-Thing Principle \cite{savage1954} inspired by the following story.

\emph{A businessman contemplates buying a certain piece of property. He considers the outcome of the next presidential election relevant. So, to clarify the matter to himself, he asks whether he would buy if he knew that the Democratic candidate were going to win, and decides that he would. Similarly, he considers whether he would buy if he knew that the Republican candidate were going to win, and again finds that he would. Seeing that he would buy in either event, he decides that he should buy, even though he does not know which event obtains, or will obtain, as we would ordinarily say}.

The Sure-Thing Principle is equivalent to the assumption that, if persons are indifferent in choosing between simple lotteries $L_1$ and $L_2$, they will also be indifferent in choosing between $L_1$ mixed with an arbitrary simple lottery $L_3$ with probability $p$ and $L_2$ mixed with $L_3$ with the same probability $p$ (\emph{independence axiom}).

Let us now consider the situation proposed by Daniel Ellsberg \cite{ellsberg1961} to point out an inconsistency with the predictions of the expected utility hypothesis and a violation of the Sure-Thing Principle. An urn contains 30 red balls and 60 balls that are either black or yellow, the latter in unknown proportion. One ball is to be drawn at random from the urn. To `bet on red' means that you will receive a prize $a$ (say, 100 \$) if you draw a red ball (`if red occurs') and a smaller amount $b$ (say, 0 \$) if you do not. Test subjects are given four options: (i) `a bet on red', (ii) `a bet on black', (iii) `a bet on red or yellow', (iv) `a bet on black or yellow', and are then presented with the choice between (i) and (ii), and the choice between (iii) and (iv). A very frequent pattern of response is that (i) is preferred to (ii), and (iv) is preferred to (iii). This violates the Sure-Thing Principle, which requires the ordering of (i) to (ii) to be preserved in (iii) and (iv). 

The contradiction above suggests that preferences of real life subjects are inconsistent with the Sure-Thing Principle. A possible explanation of this difficulty could be that people make a mistake in their choice and that the paradox is caused by an error of reasoning. We have recently studied these paradoxes, together with the existing attempts to solve them, and we instead maintain that the violation of the Sure-Thing Principle is not due to an error of reasoning but, rather, to a different type of reasoning. This reasoning is not only guided by logic but also by conceptual thinking which is structurally related to quantum mechanics \cite{aerts2009,aertsdhooghe2009}.\footnote{It should be stressed that in the simple situation considered by Ellsberg it can be argued that there is a real error of reasoning, since the probabilities can be estimated if some simple additional hypothesis about the distribution of the balls are made. We consider that the root of the Ellsberg paradox is however not an error of reasoning, but rather a different type of reasoning which is present next to classical logical reasoning, for we look at the Ellsberg situation as a metaphorical example of archetypical equivalent and more complex situations appearing in real life. Even for the Ellsberg situation in its concrete form, one could hesitate about the fact that it is always the same number of yellow balls. For such a concrete similar situation in real life that has been studied explicitly we refer to the disjunction effect, for example in the form of the Hawaii problem \cite{tverskyshafir1992,aertsczachordhooghe2011}. As we analyze it, the root there is not an error of reasoning, but a different nonclassical type of reasoning. Moreover, this quantum conceptual reasoning constitutes an essential aspect of human reasoning equally important than the aspect of human reasoning where classical logic applies.}  In particular, we have performed in \cite{aertsdhooghesozzo2011} a test of the Ellsberg paradox on a sample of real subjects. We have also asked them to explain the motivations of their choices. As a consequence of their answers, we have identified some conceptual landscapes (physical, optimistic, pessimistic, suspicion, etc.) that act as decision contexts surrounding the decision situation and influencing the subjects' choices in the Ellsberg paradox situation. We refer to \cite{aertsdhooghesozzo2011} for a complete analysis of these landscapes.
%
%(i) \emph{Physical landscape}: `an urn is filled with 30 balls that are red, and with 60 balls chosen at random from a collection of black and a collection of yellow balls'. 
%
%(ii) \emph{First choice pessimistic landscape}: `there might well be substantially fewer black balls than yellow balls in the urn, and so also substantially fewer black balls than red balls'. 
%
%(iii) \emph{First choice optimistic landscape}: `there might well be substantially more black balls than yellow balls in the urn, and so also substantially more black balls than red balls'. 
%
%(iv) \emph{Second choice pessimistic landscape}: `there might well be substantially fewer yellow balls than black balls, and so substantially fewer red plus yellow balls than black plus yellow balls, of which there are a fixed number, namely 60'.
%
%(v) \emph{Second choice optimistic landscape}: `there might well be substantially more yellow balls than black balls, and so substantially more red plus yellow balls than black plus yellow balls, of which there are a fixed number, namely 60'.
%
%(vi) \emph{Suspicion landscape}: `who knows how well the urns has been prepared, because after all, to put in 30 red balls is straightforward enough, but to pick 60 black and yellow balls is quite another thing; who knows whether this is a fair selection or somehow a biased operation, there may even have been some kind of trickery involved'.
%
%(vii) \emph{Don't Bother Me With Complications Landscape}: `if things become too complicated I'll bet on the simple situation, the one I understand well'.
%
We instead observe, as a consequence of our analysis, that it is the combined effect of all landscapes that is responsible, together with ambiguity aversion, of the experimental results collected since Ellsberg, hence of the deviations from classically expected behavior. Then, the presence of these contextual effects entails that a quantum or, better, quantum-like, formalism is needed to model the Ellsberg situation at a statistical level \cite{aertsdhooghesozzo2011}. This insight will be deepened and improved in the next sections, but we first need to introduce the notion of \emph{contextual risk}.

\section{Contextual risk and the hidden measurement formalism\label{contextualrisk}} 
Contextual risk has been formally introduced by two of us in \cite{aertssozzokavala1,aertssozzokavala2} to mathematically represent those statistical situations occurring in economics which strongly depend on context and cannot be associated with a Kolmogorovian probability model \cite{kolmogorov1933}. Let us resume the essentials of contextual risk in the following. 

Frank Knight introduced a distinction between the different kinds of uncertainty occurring in economics \cite{knight1921}, and Daniel Ellsberg inquired into the conceptual differences between them \cite{ellsberg1961}. More explicitly, Ellsberg put forward the notion of \emph{ambiguity} as an uncertainty without any well-defined probability measure modeling this uncertainty, as opposed to \emph{risk}, where such a probability measure does exist. The difference between ambiguity and risk can be grasped by considering the situation introduced by Ellsberg himself. For example, `betting on red' concerns risk, since the probability involved is known, namely 1/3 to win the bet and 2/3 to loose it. Instead, `betting on black' concerns ambiguity, since it is only known that the sum of the black and the yellow balls is 60, hence the number of black balls is not explicitly known.

Ellsberg implicitly considered classical, or Kolmogorovian, probability \cite{kolmogorov1933} in his definition of risk and ambiguity. The research on the foundations of quantum mechanics has meanwhile shown that classical probability is not the most general conceivable probabilistic framework, since it cannot model situations where context plays a crucial role. It is worth to be more explicit on this point and devote some words to it. In classical physics one can construct models which include indeterminism, e.g., statistical mechanical models. But, this indeterminism only describes the subjective lack of knowledge about the pure state in which the physical entity has been prepared. Thus, a notion of statistical, or mixed, state is introduced. The ensuing probability model satisfies the axioms of Kolmogorov (\emph{Kolmogorovian probability}). The situation is different in quantum mechanics where the probability model involved is non-Kolmogorovian \cite{accardi1982,accardifedullo1982,pitowsky1989}. One of the authors has proved that the non-Kolmogorovian nature of quantum probability can be explained as originating from a lack of knowledge about how context (in the case of quantum mechanics, measurement context) interacts with the entity that is considered, i.e. with this explanation the quantum probability is due to the presence of fluctuations in the interaction between context and entity. This idea has been elaborated and a \emph{hidden measurement formalism} has been worked out in which it has been shown that, whenever the effects of context on a (not necessarily physical) entity can be neither neglected nor predicted, then the probabilistic framework describing this situation is necessarily non-Kolmogorovian and admits either a pure (Hilbert space) quantum or a quantum-like representation \cite{aerts1986,aerts1993,aerts1995,aerts1998,aerts2002}. 

Contextual influence frequently appears in situations of risk. Let us consider, for example, the risk of having an accident. It is evident that if a person is in a context where he (she) is sitting in a chair reading the newspaper on his (her) terrace, the risk of having an accident is low if compared with the risk of getting an accident when this person is in the context of sitting in a car next to a reckless driver. In case `risk to have an accident' is considered with respect to this person, then the two mentioned contexts will have a different influence on the probabilities describing this risk. Similar examples can be found in risk management, where one has to identify, monitor and control the external factors, including accidents, natural causes and disasters, which can potentially affect given financial operations. For this reason, we have introduced in a recent paper \emph{contextual risk} to model the context dependent situations that are described in the economic literature in terms of ambiguity \cite{aertssozzokavala1,aertssozzokavala2}. The main and pragmatically relevant difference is that the notion of contextual risk can now be endowed with a probabilistic model. And, better, contextual risk can be recovered within our general hidden measurement formalism, hence it admits a non-Kolmogorovian quantum-like probability structure.

Let us now explain how the hidden measurement approach can be applied to provide a mathematical modeling for the notion of contextual risk. Economists are interested in `decisions taken by humans with respect to specific situations', such as the situation described in the Ellsberg paradox. In the process of decision making there will generally be a `cognitive influence' having its origin in the way the mind(s) of the person(s) involved in the decision making relate to the situation that is the subject of the decision making. The role played in physics by `the physical system' is played in economics by `the considered situation', the role played in physics by the measurement is played in economics by the `cognitive context', and the role played in physics by the interaction during the measurement is played in economics by the decision interaction between mind and situation. The cognitive context incorporates in principle all types of cognitive aspects that are able to influence the decision interaction. Quite obviously we generally are in a situation that there is lack of knowledge about the situation itself, but also lack of knowledge about the decision context and how it interacts with the situation. This presence of the specific double lack of knowledge is what makes contextual probability, i.e. the hidden measurement approach, apt for providing a faithful description.  
     
The way in which the hidden measurement formalism can be applied to contextual risk will be evident in the next section where an explicit hidden measurement model will be constructed for the Ellsberg paradox.

\section{A sphere model for the Ellsberg paradox\label{spheremodel}}
We elaborate in this section a macroscopic physical model that reproduces the statistical features of the decision system considered by Ellsberg. We will see that the sphere model presented here has a non-Kolmogorovian quantum-like structure, which provides a theoretical support to maintain that also the Ellsberg system should exhibit the same features \cite{aertssozzokavala1,aertssozzokavala2}.

The sphere model consists of a physical entity $S$ that is a material point particle $P$ moving on the surface of a sphere, denoted by $surf$, with center $O$ and radius 1. The unit vector ${v}$ where the particle is located on $surf$ represents the pure state $p_v$ of the entity $S$ (see Fig. 1a). For each point ${u} \in surf$, we introduce the following measurement $e_u$. We consider the diametrically opposite point $-{u} \in surf$ and install a piece of elastic of 2 units of length such that it is fixed with one of its endpoints in ${u}$ and the other endpoint in $-{u}$. Once the elastic is installed, the material particle $P$ falls from its original place ${v}$ orthogonally onto the elastic, and sticks on it (Fig. 1b). Then the elastic breaks somewhere and the particle $P$, attached to one of the two pieces of the elastic (Fig. 1c), moves to one of the two endpoints ${u}$ or $-{u}$ (Fig. 1d). Depending on whether the particle $P$ arrives at ${u}$ (as in Fig. 1) or at $-{u}$, we attribute the outcome $o_{1}^{u}$ or $o_{2}^{u}$ to $e_{u}$. The elastic installed between  ${u}$ and $-{u}$ plays the role of a (measurement) context for the entity $S$. 
\begin{figure}
\begin{center}
\includegraphics[scale=0.75]{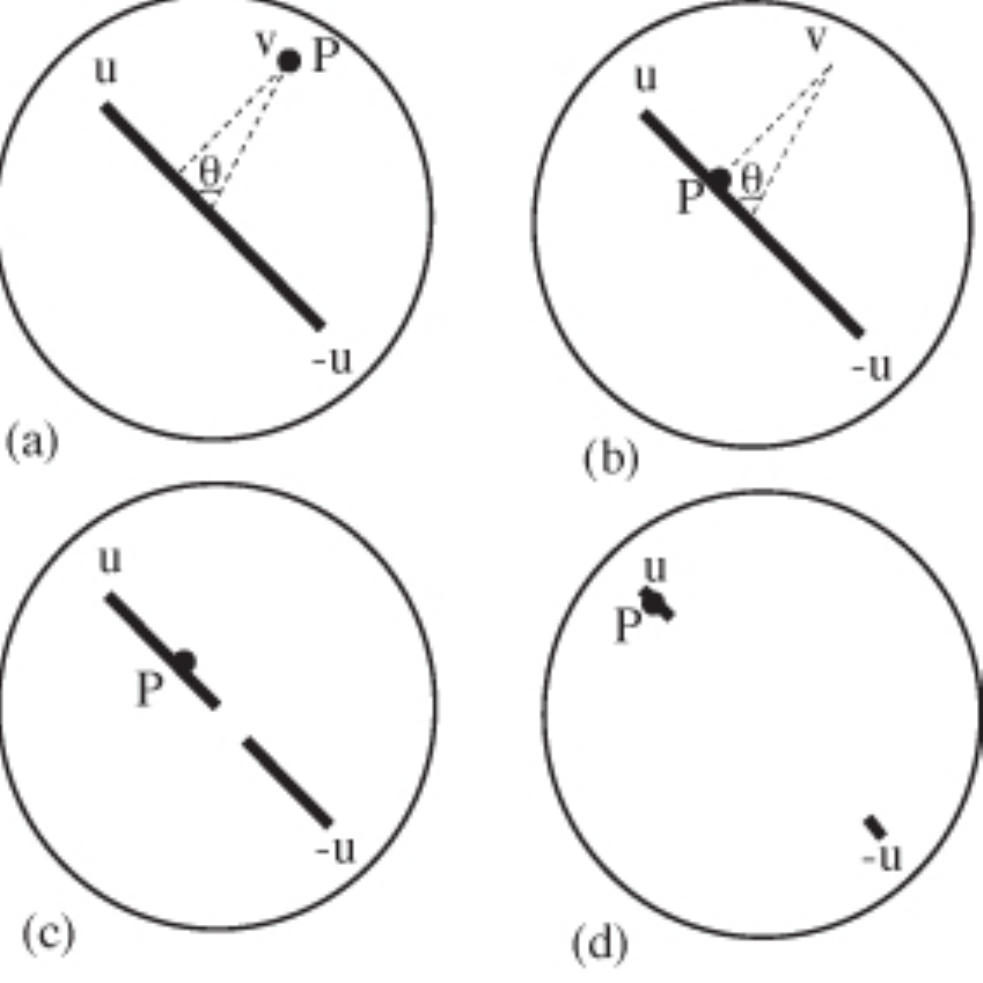}
\end{center}
\caption{A representation of the sphere model}
\end{figure}
Let us now consider elastics that break in different ways depending on their physical construction or on other environmental happenings. We can describe such a general situation by a classical probability distribution
\begin{equation}
\rho: [-u,u] \longrightarrow [0, +\infty]
\end{equation}
such that 
\begin{equation}
\int_{\Omega} \rho(x) d x  
\end{equation}
is the probability that the elastic breaks in the region $\Omega \subset [-u, u]$. We also have:
\begin{equation}
\int_{[-u,u]} \rho(x) d x =1, 
\end{equation}
which expresses the fact that the elastic always breaks during a measurement. A measurement $e_{u}$ characterized by a given $\rho$ will be called a  \emph{$\rho$-measurement} and denoted by $e_{\rho}^{u}$ in the following. A $\rho$-measurement is a hidden measurement for the entity $S$.

Let us calculate the probabilities involved in a $\rho$-measurement. The transition probabilities that the particle $P$ arrives at point ${u}$ (hence the outcome of the measurement is  $o_{1}^{u}$) and $-{u}$ (hence the outcome of the measurement is $o_{2}^{u}$) under the influence of the measurement $e_{\rho}^{u}$, are respectively given by:
\begin{eqnarray}
\mu_{\rho}(p_{u}, e_{u}, p_{v})=\int_{-1}^{{v} \cdot {u}} \rho(x) d x \nonumber \\
\mu_{\rho}(p_{-u}, e_{u}, p_{v})=\int_{{v} \cdot {u}}^{1} \rho(x) d x .
\end{eqnarray}

Let us now come the Ellsberg paradox situation, and consider an urn with 90 balls of different colors, red, black and yellow. Let us assume that the pure state $p_{v}$ represents a physical situation where the number of balls is fixed, e.g., 30 red balls, 32 black balls and 28 yellow balls (another pure state is a physical situation where the number of balls is the same, i.e. 30). Thus, different combinations of colors correspond to different sectors on the sphere. Then, for each unit vector  ${u} \in surf$, let us consider the measurement $e_{u}$ representing the decisional situation where the subject is asked to bet on a given color, red or black, and associate $e_{u}$ with the two outcomes $o_{1}^{u}$ and $o_{2}^{u}$ in such a way that, if the outcome $o_{1}^{u}$ ($o_{2}^{u}$) is obtained, this corresponds to the situation where the subject chooses to `bet on red' (`bet on black'). Moreover, let us consider different kinds of elastics, characterized by the classical probability distributions $\rho_1, \rho_2, \ldots$, one for each conceptual landscape defined in the second section. The non-uniform distributions reflect the different cognitive aspects of the decisional process. Hence, the probabilities of `betting on red' and `betting on black' under the conceptual landscape represented by the classical probability distribution $\rho_j$ are respectively given by:
\begin{eqnarray} 
\mu_{\rho_j}(p_{u}, e_{u}, p_{v})=\int_{-1}^{{v} \cdot {u}} \rho_j(x) d x \nonumber \\
\mu_{\rho_j}(p_{-u}, e_{u}, p_{v})=\int_{{v} \cdot {u}}^{1} \rho_j(x) d x . \label{pure}
\end{eqnarray}
%We note that, if we take into account the physical situation in which the urn contains 30 red balls, 30 black balls and 30 yellow balls, and locate the unit vector   representing this physical situation in the north pole of the sphere, then, for every conceptual landscape, both probabilities in Eq. (5) are equal to 1/2, which corresponds to what one actually expects.

Till now we have considered only physical situations in which the number of balls was fixed, that is, the preparation of the balls in the urn was completely known. This situation was reflected by the fact that the physical state $p_{v}$ of the entity $S$ was a pure state and the point ${v}$ located on $surf$. But we know that in the Ellsberg paradox situation only the number of red balls is known, i.e. 30 balls, while black and yellow balls are in unknown proportion. This situation can be realized in our sphere model by introducing mixed states and representing them by inner points of the sphere. For example, if the subject knows that a physical situation associated with the pure state $p_{v}$ is mixed with a physical situation associated with the pure state  $p_{-v}$ (where the point $-{v} \in surf$ is opposed to the point ${v}$) with probabilities $s \in [0,1]$ and $(1-s) \in [0,1]$, respectively, so that the state of the entity is a mixed state $p_{w}$, with ${w}=s{v}+(1-s)({-v})$, then the probabilities of `betting on red' and `betting on black' under the conceptual landscape represented by the classical probability distribution $\rho_j$ are respectively given by:
\begin{eqnarray} 
\mu_{\rho_j}(p_{u}, e_{u}, p_{w})=s \int_{-1}^{{v} \cdot {u}} \rho_j(x) d x+(1-s)\int_{-1}^{-{v} \cdot {u}} \rho_j(x) d x \nonumber \\
\mu_{\rho_j}(p_{-u}, e_{u}, p_{w})=s \int_{{v} \cdot {u}}^{1} \rho_j(x) d x +(1-s)\int_{-{v} \cdot {u}}^{1} \rho_j(x) d x. \label{mixed}
\end{eqnarray}                      
The presentation of the first part of the Ellsberg paradox is thus completed.     

It is important to observe that the probabilities in Eqs. (\ref{pure}) and (\ref{mixed}) cannot be cast into a unique Kolmogorovian scheme, which can be proven by referring to \emph{Pitowsky's polytopes}, or to \emph{Bell-like inequalities} \cite{pitowsky1989}. Furthermore, if we limit ourselves to consider uniform probability distributions $\rho_{j}$, then the probabilities in Eqs. (\ref{pure}) and (\ref{mixed}) become the standard quantum probabilities for spin measurements, since our sphere model is a model for a spin 1/2 quantum particle \cite{aerts1993,aerts1995,aerts1998}. 

In the model illustrated above we limited ourselves to consider the first part of the Ellsberg paradox, namely the situation in which a subject is asked to decide between `betting on red' and `betting on black'. A more complex model should be constructed to take into account the whole paradox. We do not accomplish this task in the present paper, for the sake of lack of space. We instead observe that our simple sphere model already shows that the Ellsberg example cannot be modeled by using classical Kolmogorovian probabilities, because of its intrinsic contextuality, and that a non-Kolmogorovian quantum-like framework is necessary. This result is relevant in our opinion and we devote the next section to explain and clarify it.

\section{Contextual risk and quantum conceptual layer\label{concl}}
The notion of contextual risk recently introduced by two of the authors \cite{aertssozzokavala1,aertssozzokavala2} to mathematically represent ambiguity situations in economics has been applied to the Ellsberg paradox. Within our hidden measurement formalism a sphere model for the Ellsberg situation has been elaborated which shows that a unique Kolmogorovian scheme is not suitable to model the experimental situation put forward by Ellsberg. Moreover, a quantum or quantum-like framework is needed because of the relevance of context in the form of conceptual landscapes in this situation. The analysis undertaken in this paper supports the hypothesis that two structured and superposed layers can be identified in human thought: a \emph{classical logical layer}, that can be modeled by using a classical Kolmogorovian probability framework, and a \emph{quantum conceptual layer}, that can instead be modeled by using the probabilistic formalism of quantum mechanics. The thought process in the latter layer is given form under the influence of the totality of the surrounding conceptual landscape, hence context effects are fundamental in this layer \cite{aerts2009,aertsdhooghe2009}.

We conclude this paper with two remarks on our approach.
%I SEPARATED THE TWO REMARKS AND MODIFIED THE TEXT, AS FOLLOWS. 

(i) The violation of the expected utility hypothesis and the Sure-Thing Principle is not (only) due to the presence of an ambiguity aversion. We argue instead that the above violation is due to the concurrence of superposed conceptual landscapes in human minds, of which some might be linked to ambiguity aversion, but other completely not. We therefore maintain that the violation of the Sure-Thing Principle should not be considered as a fallacy of human thought, as often claimed in the literature but, rather, as the proof that real subjects follow a different way of thinking than the one dictated by classical logic in some specific situations, which is context-dependent.

(ii) An explanation of the violation of the expected utility hypothesis and the Sure-Thing Principle in terms of quantum probability and quantum interference has already been presented in the literature \cite{busemeyerwangtownsend06,franco07,khrennikovhaven09,pothosbusemeyer09}. What is new in our approach is the fact that the quantum mechanical modeling is not only an elegant formal tool but, rather, it reveals the presence of an underlying quantum conceptual thought. We stress that the presence of a quantum structure in cognition and decision taking does not presuppose the existence of microscopic quantum processes in human brain. We have indeed avoided such a compelling assumption in this paper.

%\begin{theacknowledgments}
%This research was supported by Grants G.0405.08 and G.0234.08 of the Flemish Fund for Scientific Research.
%\end{theacknowledgments}

\end{document}